\documentclass{article} 

\usepackage[english]{babel} 
\usepackage{amssymb}
\usepackage{amsmath}
\usepackage{txfonts}
\usepackage{mathdots}
\usepackage[classicReIm]{kpfonts}
\usepackage[dvips]{graphicx} 

\begin{document}


\noindent 

\noindent 
\[1\]

\noindent \textbf{Dark Current Reduction of P3HT-Based Organic Photodiode Using a Ytterbium Fluoride Buffer Layer in Electron Transport}

\noindent \textbf{Seong Bin Lim, Chan Hyuk Ji, Ki Tae Kim and Se Young Oh$\boldsymbol{\mathrm{\dagger}}$}

\noindent \textbf{\textit{Department of Chemical \& Biomolecular Engineering, Sogang University, Seoul 121-742, Republic of Korea}}

\noindent \textbf{\textit{}}

\noindent \textbf{}

\noindent Photodiodes are widely used to convert lights into electrical signals. The conventional silicon (Si) based photodiodes boast high photoelectric conversion efficiency and detectivity. However, in general, inorganic-based photodiodes have low visible wavelength sensitivity due to their infrared wavelength absorption. Recently, electrical conducting polymer-based photodiodes have received significant attention due to their flexibility, low cost of production and high sensitivity of visible wavelength ranges. In the present work, we fabricated an organic photodiode (OPD) consisting of ITO/ NiOx/ P3HT:PC${}_{60}$BM/ YbF${}_{3}$/ Al. In the OPD, a yitterbium fluoride (YbF${}_{3}$) buffer layer was used as the electron transport layer. The OPD was analyzed for its optical-electrical measurements, including J-V characteristics, detectivity and dynamic characteristics. We have investigated the physical effects of the YbF${}_{3}$ buffer layer on the performance of OPD such as its carrier extraction, leakage current and ohmic characteristics. 

\noindent 



\noindent 

\noindent Email: syoh@sogang.ac.kr

\noindent Fax: \textit{:+}82-2-714-3890

\noindent \textbf{I. INTRODUCTION}

Organic photodiodes (OPD) have been extensively investigated due to their low cost of production, flexibility and light weight compared to commercially inorganic devices. Commonly, in OPDs, blends of conjugated polymers and fullerenes are widely used to create active materials such as Poly (3-hexylthiophene-2, 5-diyl) (P3HT): Phenyl-C61-butyric acid methyl ester (PC${}_{60}$BM) bulk hetero junctions (BHJ) [1-3]. However, organic materials based on devices suffer from poor performance in aspects such as stability and detectivity, because of their leakage current and oxidation. Recently, in order to reduce leakage current and oxidation, buffer layers have been introduced between the active layer and the electrodes. This layer prevents charge carrier injection and improves stability under ambient condition. Moreover, the buffer layer contributes to the parallel resistance and series resistance and${}^{ }$can improved the photo current and external quantum efficiency (E.Q.E) [4, 5]. 

Buffer layers has been developed using metal compounds, organic material and metals [5-11]. OPDs are often used to reduce the dark current density as electron/ hole blocking layers. To reduce the dark current density, buffer layers are used with materials of low work function, such as ZnO, NiO and LiF which have wide bandgaps, in organic electronic devices. Of these, LiF/Al composite cathodes are known for their superior electron injection and high electroluminescence in organic light emitting diodes (OLED), and for their reduced dark current density and high detectivity in OPDs [6, 12, 13]. In a previous study, we demonstrated the properties of ytterbium (Yb) with low work functions as an ETL in organic photovoltaic (OPV) devices and compared its performance to that of LiF [14]. In the OPV, the use of LiF as ETL was shown to have a poor performance and surface characteristics, including high series resistance and low shunt resistance [15, 16].${}^{ }$ Additionally, we inferred that OPD with LiF is less effective than with ytterbium fluoride (YbF${}_{3}$).      

In this work which is focused on reducing dark current density and responsivity, we demonstrated that OPD performance is enhanced with the use of YbF${}_{3.}$ This OPD is based on P3HT: PC${}_{60}$BM with the insertion of a YbF${}_{3}$ layer as an ETL capped by Aluminum. The results were then compared with those of LiF / Al and with those without ETL devices. The performance was measured based on optical-electrical attributes in order to analyze leakage current, detectivity and bandwidth [17-18]. To analyze the resistance components, the OPDs was measured using impedance spectroscopy in both dark and light states. To measure the response speed of our devices, we used a pulsed laser diode (LD) light source and measured the rise and fall times of the photocurrent and observed the photocurrent for a few microseconds using an oscilloscope. The performance of the resulting OPDs is herein discussed, with a particular focus on its dynamic range, detectivity and leakage current relative to the device. 

\noindent 

\noindent \textbf{II. EXPERIMENTS}

\noindent \textbf{Reagents and Materials. }Regioregular poly(3-hexylthiophene-2,5-diyl) was purchased from Aldrich Co., Ltd. [6,6]-Phenyl C61 butyricacidmethylester(PC60BM) was purchased from Nano-C Co., Ltd. The poly(-styrenesulfonate) complex was purchased from Bayer Co., Ltd. Other chemicals used were of reagent grade.\textbf{}

\noindent \textbf{}

\noindent \textbf{Fabrication of the Photodiode Cell. }Organic photodiode cells consisting of ITO/ NiO${}_{x}$/ P3HT:  PC${}_{60}$BM / YbF${}_{3}$/ Al were fabricated with an active surface area of 0.04$\mathrm{\ }{cm}^2$. Before the deposition of each layer, a patterned ITO ($\mathrm{\le}$ 20 $\Omega$/бр) glass substrate was immersed into an ultrasonic bath of deionized water, acetone and isopropyl alcohol for 15 minutes for each solvent respectively. The cleaned ITO glass substrate was then dried at 80 $\mathrm{{}^\circ}$C for 60 minutes in a vacuum oven. After applying a UV-ozone treatment for 15 minutes, Ni was thermally evaporated under a high vacuum in 10${}^{-6 }$Torr onto ITO films on glass substrates. For the oxidation of the Ni layers, this sample was oxidized by heat treatment at 400$\mathrm{{}^\circ}$C for 3 hours [19]. The sample was transferred into a glove box where a solution of  P3HT  and PC${}_{60}$BM (1:1 ratio) in dichlorobenzene (DCB) was spin-casted onto the NiO${}_{x}$ layer. Finally, YbF${}_{3}$ (2nm) and Al (100nm) were deposited by using thermal evaporation \textbf{}

\noindent \textbf{Device Characterization. }Current-voltage measurements was taken under simulated AM 1.5 solar illumination (at 100 mW/cm${}^{2}$) using a solar simulator (Newport 69920, Newport Co., Ltd., USA) and color filter. External quantum efficiency was obtained using IVUMSTATE (Spectra Pro 300i, Acton research Co., Ltd., USA). The dynamic range characteristics and impedance measurements were recorded via the Photo Response Measurement System (TNE Tech CO., LTD). 

\noindent \textbf{}

\noindent \textbf{III. Result and Discussion}

\noindent \textbf{Fig.1. (a)} shows the schematic structure of the photodiode with the YbF${}_{3}$ layer. The devices have the following structure: ITO/ NiO/ P3HT: PC${}_{60}$BM / YbF${}_{3}$/ Al. The three samples were fabricated, one each with LiF and YbF${}_{3}$ as the ETLs, and one without ETL. In \textbf{Fig. 1. (b), }the\textbf{ }J-V curves were measured in accordance with the buffer layers properties of the samples under dark conditions. The dark current density of the device with a YbF${}_{3}$ layer recorded 1.02 $\mathrm{\times }{10}^{-7}$ A/ cm${}^{2}$ at -3V, which decreased to 2.02 $\mathrm{\times }{10}^{-8}$A/ cm${}^{2}$ at -1 V.\textbf{}

\noindent \textbf{\includegraphics*[width=7.08in, height=2.12in, keepaspectratio=false]{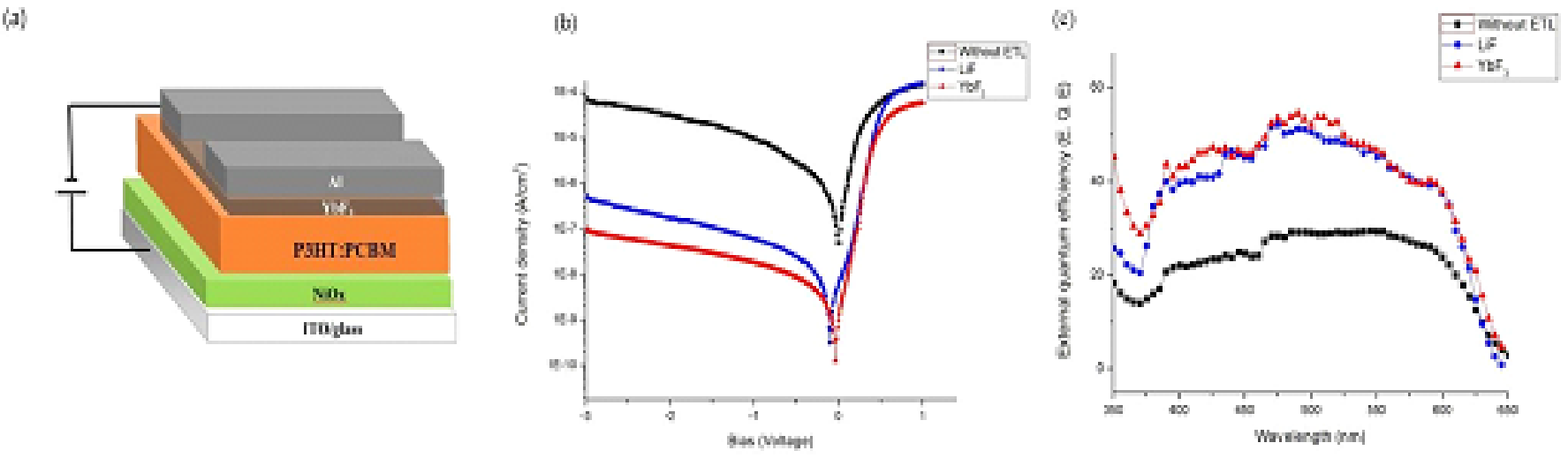}}

\noindent \textbf{Figure. 1.} (a) The cell structure of the OPD using YbF${}_{3}$, (b) the J-V characteristics of the OPD with different electron transport layers at 1sun (100 mW/cm${}^{2}$) and (c) external quantum efficiency (EQE) at 1.2mW/ cm under -1 V.

\noindent This result suggests that use of YbF${}_{3}$ significantly reduced hole injection compared to the use of LiF. The dark current density can be affected by the leakage currents at the P3HT: PC${}_{60}$BM /Al or P3HT: PC${}_{60}$BM / ETLs/ Al interface. The device without the ETL showed that contact between P3HT: PC${}_{60}$BM and Al is not suitable since the deposition of Al damages the organic surface [20]. With respect to ETL properties, YbF${}_{3}$ is suitable for covering the P3HT:PC${}_{60}$BM due to its larger optimum thickness compared to LiF and other ETL materials [21, 22]. 

\noindent With respect to illuminated conditions, the device with YbF${}_{3}$ performed the best, recording an E.Q.E of 53.73\%, while the device with LiF recorded an E.Q.E of 48.81\% at 520nm, as shown in \textbf{Fig. 1. (c)}. In the OPD, the E.Q.E value is closely related to the responsivity of the devices. The responsivity (R) is calculated using the ratio of photocurrent to incident-light intensity as shown in Equation \eqref{GrindEQ__1_},
\begin{equation} \label{GrindEQ__1_} 
\mathrm{R}\left(\mathrm{\lambdaup }\right)\mathrm{=EQE}\frac{\lambda q}{hc}\mathrm{=}\frac{I_{ph}}{L_{light}} 
\end{equation} 
where I${}_{ph}$ is the photocurrent and L${}_{light}$ is the incident light intensity [6, 18]. The responsivity ($\mathrm{R}\left(\mathrm{\lambdaup }\right)$) is used  to calculate detectivity. The detectivity (D*, Jones) indicates the photodiode ability to detect levels of incident power, as shown in Equation \eqref{GrindEQ__2_},

\noindent ${D*=(A\mathit{\Delta}f)}^{\frac{1}{2}}\bullet \frac{R\left(\lambda \right)}{I_N}\mathrm{=}\frac{R(\lambda )}{\sqrt{2qJ_{dark}}}$ [Jones, cm$\mathrm{\bullet}$${Hz}^{\frac{1}{2}}$ / W]        \eqref{GrindEQ__2_}

\noindent where A is the effective area of the diode, $\Delta$f is the bandwidth (Hz) and IN is the total noise current [6]. Here, the dark current density (J${}_{dark}$) is dominated by the shot noise (2q J${}_{dark}$) The detectivity is calculated by Equation \eqref{GrindEQ__2_} based on the measured photocurrent, dark current and incident light intensity at 520 nm for the OPDs, as shown in \textbf{Table. 1.} In the OPD, we indirectly found that the use of the YbF${}_{3}$ layer improved the performance of the resistance components in the P3HT: PC${}_{60}$BM layer compared to the use of LiF. The YbF${}_{3}$ layer was also able to cover the P3HT: PC${}_{60}$BM layer due to its large optimum thickness [15-16].${}^{ }$The use of Yb and Li composites as ETLs shows several improvements in aspects including

\noindent \includegraphics*[width=6.83in, height=2.20in, keepaspectratio=false, trim=0.00in 0.00in 0.00in 0.44in]{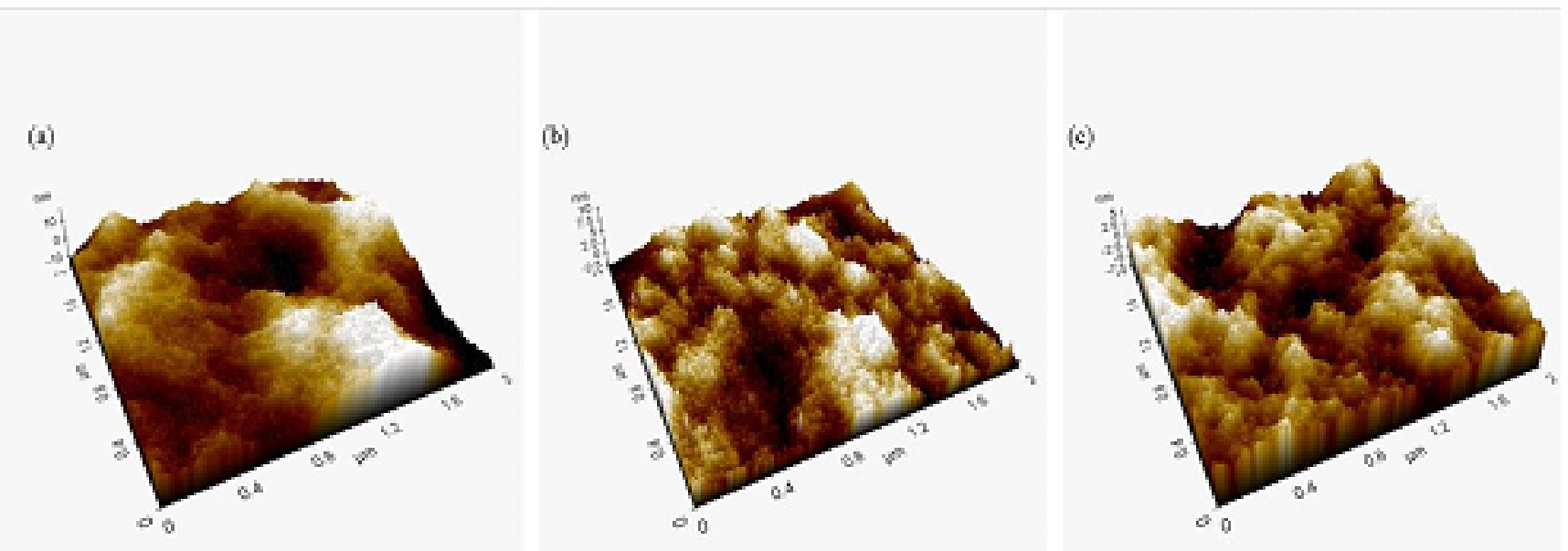}

\noindent \textbf{Fig. 2.} AFM topographic images of the surfaces of (a) ITO/ NiOx / P3HT: PC${}_{60}$BM, (b) ITO/ NiOx / P3HT: PC${}_{60}$BM / LiF and (c) ITO/ NiOx / P3HT: PC${}_{60}$BM / YbF${}_{3}$, respectively.

\noindent 

\noindent surface uniformity, contact resistance and ohmic characteristics from dipole moments. The surfaces of the ETLs on P3HT: PC${}_{60}$BM were recorded using an atomic force microscope (AFM) as shown in \textbf{Fig. 2. (a)-(c)}. From \textbf{Fig. 2.} (a), the P3HT: PC${}_{60}$BM layer showed a Rrms (roughness of root mean square) of 3.293 nm, implying poor surface uniformity. This is a result of the deposition of NiOx, as shown in \textbf{Fig. 2.} (a). In term of ETLs, the LiF and YbF${}_{3}$ layers were deposited via thermal evaporation, with optimum thicknesses of each of 0.5nm and 3nm, respectively.

\noindent 

\noindent \textbf{Table 1.} Summary of OPD parameters under -1V  

\begin{tabular}{|p{0.7in}|p{1.2in}|p{1.3in}|p{1.5in}|} \hline 
 & \textbf{J${}_{dark}$ (A/cm2)} & \textbf{E.Q.E (\%, at 520 nm)} & \textbf{Detectivity ( Jones, at 520 nm)} \\ \hline 
\textbf{Without ETL} & 9.67 x 10${}^{-6}$ & 21.09 & 3.46 x 10${}^{10}$ \\ \hline 
\textbf{LiF} & 6.44 x 10${}^{-8}$ & 48.81 & 7.15 x 10${}^{11}$ \\ \hline 
\textbf{YbF${}_{3}$} & 2.02 x 10${}^{-8}$ & 53.73 & 1.67 x 10${}^{12}$ \\ \hline 
\end{tabular}

\noindent \includegraphics*[width=6.97in, height=2.25in, keepaspectratio=false]{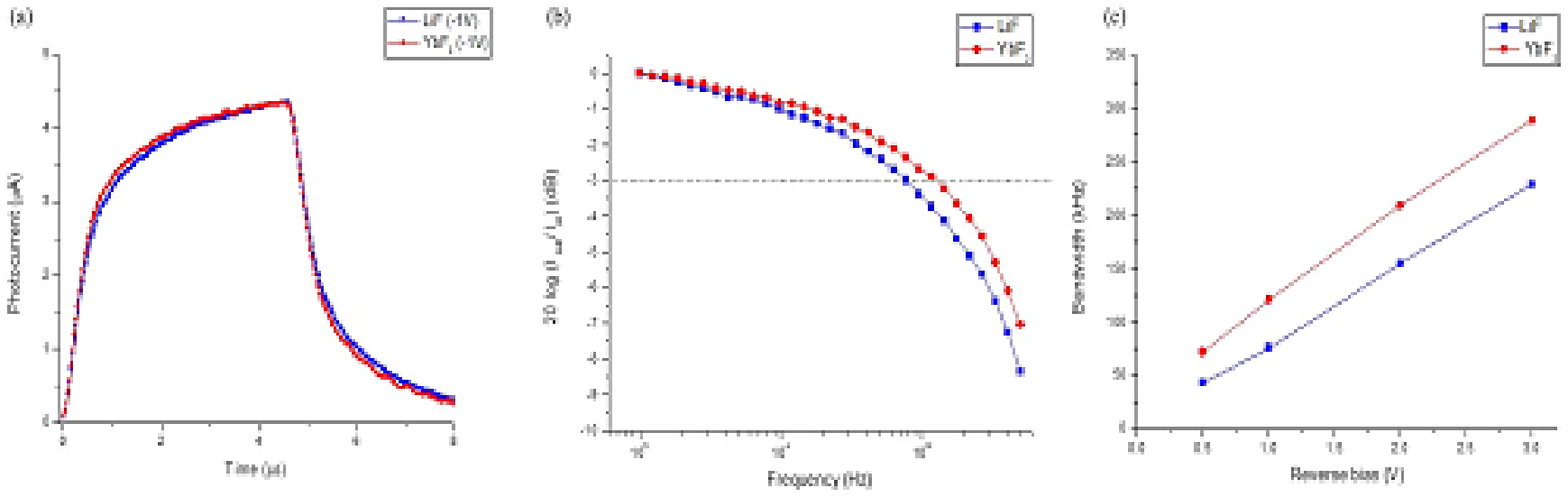}

\noindent \textbf{Fig 3. }Shows the dynamic characterization of the photocurrent response times using Laser diode with a light intensity of 650 $\muup$W/ cm${}^{2}$ at 520 nm (a) The response time under -1 V at pulsed frequency of 100 kHz are shown. (b) Cutoff frequency for OPDs under -1V (c) Bandwidth for the OPDs under reverse bias.

\noindent \textbf{}

\noindent In \textbf{Fig. 2.} (b) and (c), the Rrms values of the covered LiF and YbF${}_{3 }$were found to be 1.297 nm and 1.943 nm, respectively. The roughness of the surfaces implies that the top of P3HT: PC${}_{60}$BM can be better covered by the large optimum thickness of the YbF${}_{3}$ layer compared to the LiF layer. This is due to the above mentioned surface roughness. As such, the use of YbF${}_{3}$ as the ETL is beneficial as it decreases the contact resistance between P3HT: PC${}_{60}$BM and Al. To measure the photo response of our devices, we used a pulsed laser diode (LD) light source and measured the rise and fall times of the photocurrent on an oscilloscope. 

\noindent The shape of the response is based on the resistances and capacitance, which cause changes with respect to rise time and fall time [6]. In \textbf{Fig. 3. }(a), the response time with the YbF${}_{3 }$${}_{ }$layer showed a higher response at 1.954 $\muup$s for rise time and 2.070 $\muup$s for fall time, 2.164 $\muup$s for rise time and 2.451 $\muup$s for fall time faster than the device with LiF layer. The use of YbF${}_{3 }$as an ETL improves the surface uniformity of P3HT: PC${}_{60}$BM, which is attributed to the low contact resistance. In contrast, the use of LiF as an ETL is not recommended due to the non-uniform formation of the LiF layer, resulting in relatively high contact resistance and leakage current in the OPD. In this results, the bandwidth of the device was indirectly reflected in accordance with Equation \eqref{GrindEQ__3_}, 

\noindent \textit{??f }= 1/ (2$\piup$$\tauup$ )           \eqref{GrindEQ__3_}

\noindent where \textit{??f }is the bandwidth in Hz, and $\tauup$ is the characteristic time constant of the device [6]. \textbf{Fig. 3.} (b) and (c) showed the frequency response of the OPDs under reverse bias. In term of structure, the use of the YbF${}_{3 }$layer\textbf{ }allowed the photo excited carriers and the carriers transports to move trap-free, resulting in a faster photo response and a sharp increase in bandwidth compared to OPDs with a LiF layer. The bandwidth at -1 V of the device with the YbF${}_{3 }$${}_{ }$layer showed a higher frequency response at 120.31 kHz, 75.46 kHz higher than the device with LiF layer and indicated similar tendency under different reverse voltage. 

\noindent 

\noindent \textbf{IV. CONCLUSION}

In conclusion, we have demonstrated the performance of the organic photodiode based on P3HT: PC${}_{60}$BM. The use of YbF${}_{3}$ as an ETL reduces the dark current density by greater than one order of magnitude and improves the detectivity and responsivity by one order of magnitude compared to the use of LiF/Al as an ETL. The external quantum efficiency of the devices with YbF${}_{3}$ is 53.73\%, slightly higher than that of the device with LiF layer. This is due to the greater surface uniformity of the YbF${}_{3}$ layer. The performance of the devices in terms of detectivity and bandwidth were recorded to be 1.67 x 10${}^{12}$ and with an equally impressive cut-off frequency of 120.31 kHz at -1 V for standard device structures in OPD. The use of the YbF${}_{3}$ layer in the OPD was compatible with the commercial diodes using inorganic materials such as Si [23], InGaAs [24] and Ge [25], suggesting that OPD can be effectively applied to devices such as cameras, smart phones and image sensors, boasting improved cost of production, transparency and flexibility. 

\noindent 

\noindent \textbf{ACKNOWLEDGEMENT }

\noindent This work was supported by the Human Resources Development program (No.20114010203090) of the Korea Institute of Energy Technology Evaluation and Planning (KETEP) grant funded by the Korea

\noindent government Ministry of Trade, Industry and Energy. This research was supported by Basic Science Research Program through the National Research Foundation of Korea (NRF) funded by the Ministry of Education(NRF-2014R1A1A2055777).

\noindent \textbf{REFERENCES}

\noindent [1] P. Peumans, V. Bulovic, and S. R. Forrest, Appl. Phys. Lett. \textbf{76}, 3855 (2000).

\noindent [2] G. G. Belmonte, A. Munar, E. M. Barea, J. Bisquert, I. Ugarte, R. Pacios, Organic Electronics.  \textbf{9}, 847-851 (2008).

\noindent [3] D. Baierl, B. Fabel, P. Lugli, G. Scarpa, Organic Electronics.  \textbf{9}, 1669-1673 (2011). 

\noindent [4] O. M. Ntwaeaborwa, R. Zhou, L. Qian, S. S. Pitale, J. Xue, H. C. Swart, P. H. Holloway, Physica  B \textbf{407}, 1631--1633 (2012)

\noindent [5] P. R. Brown, R. R. Lunt, N. Zhao, T. P. Osedach, D. D. Wanger, L. Y. Chang, M. G. Bawendi and V. Bulovi, Nano Lett. \textbf{11}, 2955--2961 (2011)

\noindent [6] B. Arredondo, C. de Dios, R. Vergaz, A.R. Criado, B. Romero, B. Zimmermann, U. W\"{u}rfel, Organic Electronics. \textbf{14}, 2484--2490 (2013)

\noindent [7] M. D. Irwin, D. B. Buchholz, Al. W. Hains, R. P. H. Chang and T. J. Marks, PNAS. \textbf{8}, 105

(2008)

\noindent [8] H. S. Kim, K. T. Lee, C. Zhao, L. J. Guo, J. Kanick, Organic Electronics. \textbf{20}, 103--111 (2015) 

\noindent [9] J. Kettle, S. W. Chang, M. Horie, IEEE Sensors Journal. \textbf{15}, 3221-3224 (2015) 

\noindent [10] D. M. Im , H. U. Moon , M. C. Shin , J. H. Kim , and S. H. Yoo, Adv. Mater\textit{. }\textbf{23}, 644--648 (2011)

\noindent [11] M. Punke, S. Valouch, S. W. Kettlitz, N. Christ, C. G\"{a}rtner, M. Gerken, and U. Lemmer, APPLIED PHYSICS LETTERS. \textbf{91}, 071118 (2007)

\noindent [12] L. S. Hung, C. W. Tang, M. G. Mason, P. Raychaudhuri and J. Madathil, APPLIED PHYSICS LETTERS, 78, 544 (2001)

\noindent [13] D. Y. Kondakov, JOURNAL OF APPLIED PHYSICS. \textbf{99}, 024901 (2006)

\noindent [14] G. M. Kim, I. S. Oh, A. N. Lee and S. Y. Oh, J. Mater. Chem. A. \textbf{2}, 10131--10136 (2014)

\noindent [15] Y. Li, L. Duan, Q. Liu, R. Zhang, D. Zhang, L. Wang, J. Qiao, Y. Qiu, Applied Surface Science \textbf{254}, 7223--7226 (2008)

\noindent [16] J. R. Lian, X. Luo, W. Chen, S. X. Su, H. F. Zhao, S. Y. Liu, G. W. Xu, F. F. Niu, P, J. Zeng, Chin. Phys. Lett. \textbf{31}, 118501 (2014)

\noindent [17] P. E. Keivanidis, S. H Khong, P. K. H. Ho, N. C. Greenham and R. H. Friend, APPLIED PHYSICS LETTERS, \textbf{94}, 173303 (2009) 

\noindent [18] X. Gong, M. Tong, Y. Xia, W. Cai, J. S. Moon,Y. Cao, G. Yu, C. L. Shieh, B. Nilsson, A. J. Heeger, SCIENCE. \textbf{325}, 1665 (2009)

\noindent [19] S. H. Chang, S. C. Chae, S. B. Lee, C. Liu, T. W. Noh, J. S. Lee, B. Kahng, J. H. Jang, M. Y. Kim, D. W. Kim and C. U. Jung, APPLIED PHYSICS LETTERS. \textbf{92}, 183507 \_2008\_

\noindent [20] M. Vogel, S. Doka, C. Breyer, M. C. Lux-Steiner and K. Fostiropoulosa, APPLIED PHYSICS LETTERS. \textbf{89}, 163501 (2006)

\noindent [21] S.J. Kang, D.S. Park, S.Y. Kim, C.N. Whang, K. Jeong, S. Im, Appl. Phys. Lett. \textbf{81}, 2581(2002)

\noindent [22] C.I. Wu, C.T. Lin, Y.H. Chen, M.H. Chen, Y.J. Lu, C.C. Wu, Appl. Phys. Lett. 88, 152104 (2006)

\noindent [23] G. Konstantos, J. Clifford, L. Levina AND E. H. Sargent, nature photonics. \textbf{1}, 531-534 (2007)

\noindent [24] J. Jiang, S. Tsao, T. O Sullivan, W. Zhang, H. Lim, T. Sills, K. Mi and M. Razeghia, APPLIED PHYSICS LETTERS. \textbf{84}, 2166 (2004);

\noindent [25] C. Miesner, K. Brunner, G. Abstreiter,  Infrared Physics \& Technology. \textbf{42}, 461-465 (2001)

\noindent

\end{document}